\documentclass[prd,superscriptaddress,floatfix,showpacs,10pt,letterpaper]{revtex4}
\usepackage{graphicx} \usepackage{subfigure} \usepackage{amssymb}
\usepackage{verbatim} \usepackage{amsmath} \usepackage{amscd}
\usepackage{amssymb} 
\usepackage{epsfig}     

\input{epsf}
\usepackage{epsfig}
\usepackage{amssymb}
\usepackage{amsfonts}
\usepackage{amsbsy}
\usepackage{amsmath}

\usepackage{epstopdf}
\usepackage{graphics}      
\usepackage{graphicx}      

\begin{document}
\title{Tunneling between single and multi-centered black hole configurations}

\author{Hyeyoun Chung}
\email{hyeyoun@physics.harvard.edu}
\affiliation{Center for the Fundamental Laws of Nature, Harvard University,\\ 17 Oxford St., Cambridge, MA 02138, USA}
\date{\today}
\begin{abstract}We find a gravitational instanton that connects an initial state corresponding to a single-centered extremal Reissner-Nordstrom (ERN) black hole configuration, to a final state corresponding to a multi-centered configuration. This instanton is interpreted as describing quantum tunneling between the two different black hole solutions. We evaluate the Euclidean action for this instanton and find that the amplitude for the tunneling process is equal to half the difference in entropy between the initial and final configurations.\end{abstract}
 \maketitle 

\section{Introduction}\label{sec-Intro}

Instantons are solutions of the Euclidean equations of motion. As tunneling processes can be considered as propagation in Euclidean time, an instanton can describe tunneling transitions that are classically forbidden: for example, gravitational instantons, which are Riemannian solutions of the Einstein field equations, can describe changes in the topology of space\cite{Wheeler}. In the semiclassical WKB approximation, the tunneling amplitude is given by the value of the Euclidean action for the instanton.

Gravitational instantons have been found that describe the pair production of Wheeler wormholes\cite{Garfinkle}, and the fragmentation of a single charged $AdS_2 \times S_2$ universe into multiple, completely disconnected $AdS_2\times S_2$ universes\cite{Brill, AdSFrag}. But so far, an instanton has not been found that describes a transition between a single-centered black hole configuration to a \textit{connected} multi-centered black hole configuration.

\begin{figure}\includegraphics[width=150pt]{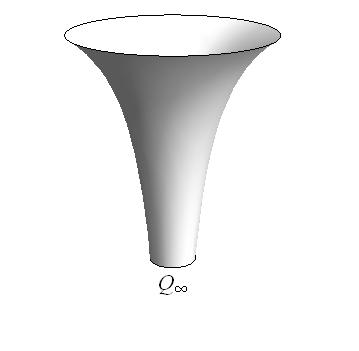}
\caption{The single-centered configuration, which is simply an $AdS_2\times S_2$ Bertotti-Robinson universe with charge $Q_\infty$.}
\label{fig-OneThroat}
\end{figure}

\begin{figure}\includegraphics[width=150pt]{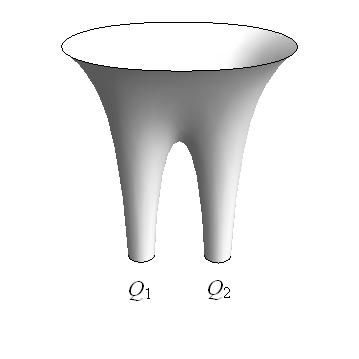}
\caption{A two-centered configuration, which consists of an encapsulating BR universe (i.e. an $AdS_2\times S_2$ throat) containing two extremal Reissner-Nordstrom black holes of charge $Q_1$ and $Q_2$.}
\label{fig-TwoThroat}
\end{figure}

In this work we describe such an instanton, where the single-centered configuration is a charged Bertotti-Robinson (BR) universe (corresponding to the $AdS_2\times S_2$ near-horizon region of an extremal Reissner-Nordstrom black hole)\cite{BertRob}, as shown in Figure \ref{fig-OneThroat}, and the multi-centered configuration consists of a set of extremal Reissner-Nordstrom (ERN) black holes placed at arbitrary locations in an encapsulating BR universe, as shown in Figure \ref{fig-TwoThroat} (the multi-centered configuration can also be interpreted as a single encapsulating $AdS_2\times S_2$ throat that divides into multiple branches as we move deeper into the throat.) The total charge of the multi-centered solution is equal to the charge of the single-centered solution. Our instanton differs from the instanton describing the pair production of black holes\cite{Garfinkle}, as it describes the splitting of an existing black hole throat. It also differs from the well-known Brill instanton\cite{Brill} describing the fragmentation of a single $AdS_2\times S_2$ throat into several completely disconnected $AdS_2\times S_2$ throats. Brill's instanton describes vacuum tunneling between a single $AdS_2\times S_2$ universe, and a set of multiple, disconnected, disjoint $AdS_2\times S_2$ universes. Our instanton, on the other hand, describes vacuum tunneling between a single $AdS_2\times S_2$ throat (corresponding to the black hole throat of a magnetically charged ERN black hole), and a branching $AdS_2\times S_2$ throat (corresponding to the black hole throat of a magnetically charged ERN black hole that has split into several throats, the total charge of which is equal to the single throat configuration.) So the Brill instanton can be interpreted as describing the \textit{fragmentation} of a single $AdS_2\times S_2$ universe. Our instanton can be interpreted as describing the \textit{splitting} of a single $AdS_2\times S_2$ throat, which nevertheless remains in a single piece at spatial infinity. Thus, our instanton describes the splitting of an ERN black hole into multiple black holes, within the \textit{same} universe: in contrast to the Brill instanton, which describes the fragmentation of one universe into several.

This instanton is analogous to the well-known instanton solution for the symmetric double well in one-dimensional quantum mechanics, as it connects two degenerate vacua. Thus, rather than describing a decay process, this instanton describes quantum \textit{mixing} between two vacuum states. The true ground state will therefore be a quantum superposition of all such configurations, where each configuration satisfies the condition that the sum of the charges of all the black holes is equal to the total charge of the single-centered configuration. Calculating the tunneling amplitude between the vacua, we find that it is equal to half the difference in entropy between the two configurations. This result agrees with the expectations of \cite{SUSYCC}, where it was suggested that the splitting of ERN black holes could be exponentially suppressed by a tunneling amplitude proportional to the change in entropy before and after the splitting.

This paper is structured as follows. In Section \ref{sec-ScSol} we describe the multi-centered black hole solutions. In Section \ref{sec-Instanton} we find the instanton solution corresponding to tunneling between single and multi-centered configurations. In Section \ref{sec-Action} we evaluate the Euclidean action for the instanton to find the tunneling amplitude. We conclude in Section \ref{sec-Conc}.

\section{The Black Hole Solution}\label{sec-ScSol}

In this section we introduce the black hole solutions that describe the two states connected by our instanton. These are solutions to the Einstein-Maxwell equations that describe magnetically charged black holes. In a spacetime with coordinates $(t,\vec{x})$, the metric and the electromagnetic field strengths are given by:
\begin{align}
ds^2 &= -H^{-2}dt^2 + H^2d\vec{x}^2\\
\star F &= dt \wedge dH^{-1},
\end{align}
where $H$ is a harmonic function that satisfies
\begin{align}
\nabla^2 H = 0,
\end{align}
where $\nabla^2$ is the Laplacian on flat $\mathbb{R}^3$. This solution describes a Bertotti-Robinson (BR) type universe\cite{BertRob} containing a number of ERN black holes. The black holes may be located at arbitrary coordinate locations $\vec{x}_a$. If there are $N$ black holes of charges $Q_1,\dots, Q_N$, then the function $H$ has the form:
\begin{align}
H = \sum_{a=1}^N \frac{Q_a}{|\vec{x} - \vec{x}_a|}.
\end{align}
We denote the total charge by $Q_\infty \equiv Q_1 + \dots + Q_N$. This solution has a tree-like geometry that is asymptotically $AdS_2\times S_2$ at large radius, but then branches into smaller $AdS_2\times S_2$ regions at the ERN black holes. The special case when $N=1$ corresponds to a single-centered black hole solution, which is simply an $AdS_2\times S_2$ space of charge $Q_\infty = Q_1$.

These solutions can also be derived by taking the limit $L_p\to 0$ (where $L_p$ is the Planck length) of asymptotically flat multi-centered ERN black hole solutions with separations of order $L_p^2$ between the black holes\cite{AdSFrag}. As $L_p \to 0$, the asymptotically flat region decouples and we are left with an encapsulating throat of charge $Q_\infty$ which splits into $N$ throats of charge $Q_1,\dots, Q_N$.

\section{The Gravitational Instanton}\label{sec-Instanton}

We want to consider a tunneling process in which a single-centered black hole configuration of charge $Q_\infty$ splits into a multi-centered black hole configuration of charges $Q_1, \dots, Q_N$ such that $Q_\infty = Q_1 + \dots + Q_N$. This process should be described by a gravitational instanton, which is a solution to the Euclidean equations of motion. In order to describe a valid tunneling process\cite{GibbonsHartle}, the Euclidean action of the instanton should be finite and real. Moreover, we should be able to define a Euclidean time coordinate $\tau$ such that slicing up the spacetime into hypersurfaces of constant $\tau$ takes us from an initial hypersurface $\Sigma_i$ at $\tau = -\infty$ to a final hypersurface $\Sigma_f$ at $\tau = \infty$. The induced metrics on $\Sigma_i$ and $\Sigma_f$ should be real, and correspond to spatial slices of the Lorentzian spacetimes that we want to connect via a tunneling trajectory (in this case, the single-centered and multi-centered solutions.) In order to allow the Euclidean solution to be joined smoothly to the Lorentzian solutions across $\Sigma_i$ and $\Sigma_f$, the extrinsic curvature $K_{ij}$ should vanish on both surfaces.

Our instanton satisfies all of these conditions\footnote{As mentioned in the introduction, our instanton differs from previously discovered instantons connecting single-centered to multi-centered black hole configurations as our instanton connects a single-centered configuration to a \textit{connected} multi-centered configuration, rather than several completely disconnected spaces. In addition to this fundamental difference, there are also differences between in the explicit technical construction of the instanton. Brill's instanton has a limitation, in that Brill does not define initial and final surfaces that can be connected to Lorentzian spacetimes. Instead he merely defines ``asymptotic regions'' at past infinity and future infinity for Euclidean time, that approach the single $AdS_2\times S_2$ throat and multiple disconnected $AdS_2\times S_2$ throats respectively. However, it is not then clear how these asymptotic regions can join smoothly onto spacelike hypersurfaces of a Lorentzian spacetime. Maldacena, Michelson, and Strominger (MMS)\cite{AdSFrag} attempted to remedy this situation by defining an instanton using different coordinates, with initial and final surfaces of zero extrinsic curvature that corresponded to spacelike slices of a single AdS space, and multiple disconnected AdS spaces respectively. However, this result still had several differences from our work. Firstly, of course, the final surface of the MMS instanton corresponds to completely disconnected $AdS_2\times S_2$ universes, unlike our work where the multiple $AdS_2\times S_2$ throats remain connected at the ``top'' of the throat. Secondly, the initial and final surfaces of the MMS instanton correspond to spatial slices of \textit{global} $AdS_2\times S_2$ spacetime, not Poincare $AdS_2\times S_2$, as was the case with Brill, and as is the case in this paper. This is an important distinction when it comes to interpreting the instanton as the splitting of the throat of an ERN black hole. Secondly, MMS assumed (but did not show) that the value of the Euclidean action for their instanton would be the same as the value Brill found. This might not necessarily be the case, given that MMS use a different set of coordinates from Brill.}. We define initial and final surfaces with zero extrinsic curvature. The initial surface is reached at $\tau = -\infty$ and the final surface at $\tau = \infty$, where $\tau$ is a suitably defined notion of Euclidean time. The initial surface is diffeomorphic to the spatial slice of a Lorentzian Poincare $AdS_2\times S_2$ throat, and the final surface is diffeomorphic to the spatial slice of a Lorentzian spacetime that is the aforementioned Poincare $AdS_2 \times S_2$ space that splits into multiple $AdS_2 \times S_2$ spaces further down the throat, but remain joined at the ``top'' of the throat. It is important that we are considering Poincare rather than global $AdS_2\times S_2$ space times, as we want to consider the splitting of black hole throats, and the throat of an ERN black hole is given by a Poincare AdS geometry.

Since we are dealing with Poincare $AdS_2\times S_2$ throats at all times, the initial and final surfaces can be thought of as spatial slices through the black hole throats of two different black hole configurations. The initial surface corresponds to a spatial slice through a single ERN black hole throat. The final surface corresponds to a spatial slice through the throat of an ERN black hole that has split into multiple ERN black holes further down the throat, but remain connected at the top of the original single throat so that the entire configuration looks like a single ERN black hole when viewed from far away.

\begin{figure}\includegraphics[width=250pt]{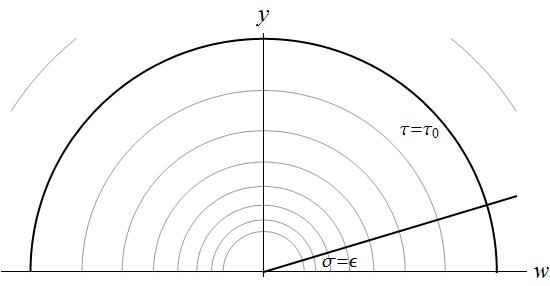}
\caption{The relationship between the coordinates $(w,y)$ and $(\tau,\sigma)$. The semicircles represent hypersurfaces of constant $\tau$, and $\sigma$ is the angular coordinate.}
\label{fig-YCutoff}
\end{figure}

In order to find the gravitational instanton, we first analytically continue the time coordinate to $w = -it$ to obtain a solution to the Euclidean equations of motion:
\begin{align}
ds^2 &= H^{-2}dw^2 + H^2d\vec{x}^2\label{eq-EucMet}\\
\star F &= -dw \wedge dH^{-1},
\end{align}
We then define the coordinate\cite{AdSFrag}:
\begin{align}
y = \left (\sum_{a=1}^N \frac{Q_a}{\sqrt{|\vec{x} - \vec{x}_a|}} \right )^2
\end{align}
Finally, we can define the coodinates $\tau$ and $\sigma$ (where in the end, we will take $\tau$ to be our Euclidean time coordinate):
\begin{align}
\tau &= \frac{1}{2}\log (w^2 + y^2)\\
\sigma &= \arctan{\frac{y}{w}}
\end{align}
The relation between the coordinates $(w,y)$ and $(\tau,\sigma)$ are shown in Figure \ref{fig-YCutoff}, where the semicircles represent the hypersurfaces of constant $\tau$, and $\sigma$ is the angular coordinate. We let $\tau$ take the range of values $-\tau_0 < \tau < \tau_0$ for some $\tau_0$, and we take $\sigma$ to cover the range $\epsilon < \sigma < \pi/2$ for some infinitesimal $\epsilon > 0$. We have regulated the lower limit of $\sigma$ as $\sigma = 0$ corresponds to spatial infinity. In the end we will first take the limit $\epsilon \to 0$, then $\tau_0\to\infty$.

We can now define the initial and final surfaces of the instanton. The initial surface $\Sigma_i$ is the hypersurface with $\tau = -\tau_0$ and $\epsilon < \sigma < \pi/2$, and the final surface $\Sigma_f$ is the hypersurface with $\tau = \tau_0$ and $\epsilon < \sigma < \pi/2$. We now have to show that the induced metric on these surfaces corresponds to spatial slices of the desired initial and final black hole configurations as we take the limit $\epsilon\to 0$, $\tau_0\to\infty$, and that the extrinsic curvatures on these surfaces is zero.

On the initial surface $\Sigma_i$, we have $w^2 + y^2 \to 0$ as $\tau_0\to\infty$. So we have
\begin{align}
y \to \frac{Q_\infty^2}{|\vec{x}|}
\end{align}
And the metric (\ref{eq-EucMet}) becomes:
\begin{align}
ds^2 &= Q_\infty^2\left (\frac{dw^2 + dy^2}{y^2}+ d\Omega^2\right ) \\
&= Q_\infty^2\left (\frac{ d\sigma^2}{\sin^2\sigma} + d\Omega^2 \right)
\end{align}
Explicit computation shows that the extrinsic curvature on this surface is zero. We can glue this to a spatial slice of Euclidean Poincare $AdS_2\times S_2$ with charge $Q_\infty$\cite{PiolineTroost}, which is the initial spacetime we want, by applying the coordinate transformation:
\begin{align}
Y &= \frac{2e^{\tau_0}\sin\sigma}{\cos\sigma + \cosh(\tau + \tau_0)}\\
W &= \frac{2e^{\tau_0}\sinh(\tau + \tau_0)}{\cos\sigma + \cosh(\tau + \tau_0)}
\end{align}
The metric on $\Sigma_i$ in these coordinates is:
\begin{align}
ds^2 &= Q_\infty^2\left (\frac{dY^2}{Y^2}+ d\Omega^2\right)
\end{align}
The coordinate $Y$ covers the range $\epsilon e^{\tau_0} < Y < e^{\tau_0}$. If we first take the cutoff $\epsilon\to 0$, then take the limit $\tau_0\to\infty$, this covers the entire $AdS_2\times S_2$ space. We can therefore join $\Sigma_i$ smoothly onto a spatial slice of a Lorentzian single-centered $AdS_2\times S_2$ solution with charge $Q_\infty$.

\begin{figure}\includegraphics[width=250pt]{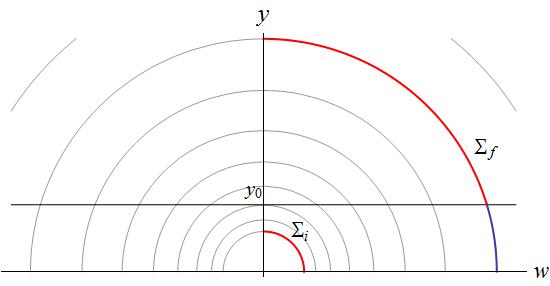}
\caption{(Color online) The initial and final surfaces of the gravitational instanton. The initial surface $\Sigma_i$ is given by the hypersurface $\tau = -\tau_0$, and the final surface $\Sigma_f$ is given by the hypersurface $\tau = \tau_0$. In order to define the coordinate transformation that allows $\Sigma_f$ to be glued to a spatial slice of the multi-centered solution, we define two regions on $\Sigma_f$: the blue region corresponds to $y < y_0 + \delta$, and the red region corresponds to $y > y_0 - \delta$, for some small fixed $\delta$.}
\label{fig-Surfaces}
\end{figure}

The final surface $\Sigma_f$ is slightly more complicated. We first choose a cutoff $y=y_0$ for the $y$-coordinate. Then for the range of coordinates $y<y_0 + \delta$ on $\Sigma_f$ for some fixed, small $\delta$, we define the coordinates:
\begin{align}\label{eq-Sigmaf1}
Y &= y\\
W &= w\nonumber
\end{align}
This section of $\Sigma_f$ corresponds to the blue segment of the hypersurface in Figure \ref{fig-Surfaces}. For fixed $y_0$, as $\tau_0 \to \infty$, we find that $dw\to 0$ in this region. Specifically, the time-component of the metric is $H^{-2}dw^2$, and we have:
\begin{align}
H^{-1}dw &\to H^{-1}(e^{\tau}d\tau - e^{\tau}\sin\sigma d\sigma)\\
&\to -y_0H^{-1}d\sigma = O(e^{-\tau_0})\\
&\to 0
\end{align}
as $\tau_0\to\infty$, for $y < y_0 + \delta$. Thus the metric on this section of $\Sigma_f$ is simply the Poincare spatial slice of the full multi-centered solution for the range of coordinates $\epsilon e^{\tau_0} < y < y_0 + \delta$. Explicit computation shows that the extrinsic curvature on this section vanishes.

For the range of coordinates $y > y_0 - \delta$ (corresponding to the red segment of the hypersurface $\Sigma_f$ in Figure \ref{fig-Surfaces}), if we choose a large enough cutoff $y=y_0$, then we are on one of the throats of the multi-centered solution, as we are in the region of large $y$. Thus we have
\begin{align}
y \to \frac{Q_a^2}{|\vec{x}|}
\end{align}
for one of the charges $Q_a$, where we have redefined the coordinate $\vec{x}$ so that the origin is centered at $Q_a$. The metric approaches
\begin{align}
ds^2 &= Q_a^2\left (\frac{dw^2 + dy^2}{y^2}+ d\Omega^2\right ) \\
&= Q_a^2\left (\frac{ d\sigma^2}{\sin^2\sigma} + d\Omega^2 \right)
\end{align}
as $y_0\to\infty$. Explicit computation shows that the extrinsic curvature vanishes on this section of $\Sigma_f$. In this region, we apply the coordinate transformation:
\begin{align}\label{eq-Sigmaf2}
Y &= \frac{2e^{\tau}\sin\sigma}{\cos\sigma + \cosh(\tau - \tau_0)}\\
W &= \frac{2e^{\tau}\sinh(\tau - \tau_0)}{\cos\sigma + \cosh(\tau - \tau_0)}\nonumber
\end{align}
Note that in the overlap region, $y_0 - \delta < y < y_0 + \delta$, the definitions of the two sets of coordinates (\ref{eq-Sigmaf1}) and (\ref{eq-Sigmaf2}) agree in the limit $\tau_0 \to \infty$ for any fixed cutoff $y_0$.

The metric on this segment $y > y_0 - \delta$ of $\Sigma_f$ is, for each throat:
\begin{align}
ds^2 &= Q_a^2\left (\frac{dY^2}{Y^2}+ d\Omega^2\right)
\end{align}
This is the Poincare spatial slice of an $AdS_2\times S_2$ throat of charge $Q_a$. 

Thus, we find that the blue segment of $\Sigma_f$ (corresponding to $y < y_0 + \delta$), can be glued to a spatial slice of a multi-centered solution cut off at $y=y_0$, while the red segment consists of the spatial slices of multiple $AdS_2\times S_2$ throats attached to the multi-centered solution at the cutoff. Note that the cutoff $y_0$ may be taken to be arbitrarily large. As with the initial surface $\Sigma_i$, the coordinate $Y$ covers the range $\epsilon e^{\tau_0} < Y < e^{\tau_0}$. If we first take $\epsilon\to 0$, then take $\tau_0\to\infty$, this covers the entire multi-centered solution. So $\Sigma_f$ may be glued to the spatial slice of a multi-centered solution.

\section{The Value of the Euclidean Action}\label{sec-Action}

We can now evaluate the Euclidean action for the instanton. The action is given by\cite{HawkingIsrael}:
\begin{align}
-16\pi I = \int (R - F_{\mu\nu}F^{\mu\nu})\sqrt{g}d^4x + 2\int K\sqrt{h}d^3 x + C[h_{ij}]
\end{align}
where $h_{ij}$ is the induced metric on the boundary, $K$ is the trace of the second fundamental form of the boundary, and $C[h_{ij}]$ is a term that depends solely on the induced metric at the boundary. The action can be converted into a pure boundary term\cite{Brill}:
\begin{align}\label{eq-ActionBd}
-8\pi I = \int (K + C_\mu \star F^{\mu\nu}n_\nu)\sqrt{h}d^3 x + C[h_{ij}]
\end{align}
where $C_\mu$ is a vector potential defined by:
\begin{align}
\star F = dC.
\end{align}

The spacetime is bounded by the following surfaces: the constant $\tau$ hypersurfaces $\tau = \pm\tau_0$, $\epsilon < \sigma < \pi/2$, the constant $\sigma$ hypersurface $\sigma = \pi/2$, $-\tau_0 < \tau < \tau_0$, and the constant $\sigma$ hypersurface $\sigma = \epsilon$, $-\tau_0 < \tau < \tau_0$.

On the first three hypersurfaces $\tau = \pm\tau_0$, $\epsilon < \sigma < \pi/2$ and $\sigma = \pi/2$, $-\tau_0 < \tau < \tau_0$, both the extrinsic curvature term and the electromagnetic term in (\ref{eq-ActionBd}) vanish, and thus do not contribute to the Euclidean action. The constant $\sigma$ hypersurface $\sigma = \epsilon$, $-\tau_0 < \tau < \tau_0$ corresponds to the hypersurface $|\vec{x}|=e^{-\tau_0}/\epsilon$, $e^{-\tau_0} < w < e^{\tau_0}$. On this surface, we find that
\begin{align}
K &= \frac{1}{Q_\infty}\\
C_\mu \star F^{\mu\nu}n_\nu &= -\frac{1}{Q_\infty},
\end{align}
as $\epsilon\to 0$, so the two terms cancel, and this hypersurface also does not contribute to the Euclidean action.

Thus the only non-zero contributions to the Euclidean action come from the ``edges'' of the spacetime, at $\tau = \pm\tau_0$, $\sigma = \epsilon$, and $\tau = \pm\tau_0$, $\sigma = \pi/2$. The contribution from the ``edges'' can be calculated using the results in \cite{Hayward}: the contribution from an edge formed by two boundaries with spacelike normals $n_0, n_1$ (which is the case here, in Euclidean spacetime) is given by:
\begin{align}
\int_{edge} -i\eta \sigma^{1/2}d^2x = -i\eta A_a
\end{align}
$A_a$ is the area of the edge $|\vec{x} - \vec{x}_a| \to 0$, which is $4\pi Q_a^2$. The factor of $-i$ comes from Wick-rotation, and we have defined
\begin{align}
\eta\equiv \mathrm{arccosh}(-n_0\cdot n_1)
\end{align}
In this case we have:
\begin{align}
\mathrm{arccosh}(-n_0\cdot n_1) &\to \mathrm{arccosh}\left(0\right)
\end{align}
at all the edges. Note that the directions of the normals (i.e. whether they are inward or outward-pointing) does not matter, since we can take the branch of arccosh such that
\begin{align}
\mathrm{arccosh} \left(0\right) = \frac{i\pi}{2}
\end{align}
whether 0 is approached from above or below. Thus each edge will give a contribution of the same sign.

As with the contribution from the mantle surfaces, we want to normalize the Euclidean action so that it is zero when evaluated on a single-centered solution of total charge $Q_\infty = Q_1$. Without normalizing, the edge terms for such a solution add up to:
\begin{align}
\frac{1}{8\pi}(\frac{\pi}{2} A_1 + \frac{3\pi}{2} A_\infty) = \frac{1}{4} A_\infty,
\end{align}
where the first term comes from the edge $\tau = \tau_0$, $\sigma = \pi/2$, and the second term comes from the edges $\tau = -\tau_0$, $\sigma = \pi/2$, and $\tau = \pm\tau_0$, $\sigma = \epsilon$. So we must subtract $\frac{1}{4} A_\infty$ to obtain the correct normalization (this can be done by setting $C[h_{ij}] = -\frac{1}{4} A_\infty$.) This means that the total contribution from the edge terms is:
\begin{align}
-\frac{1}{8\pi}\left(\sum_{a=1}^N \frac{\pi}{2} A_a + \frac{3\pi}{2} A_\infty - 2\pi A_\infty \right)&= -\frac{1}{16}\sum_{a=1}^N A_a +\frac{1}{16} A_\infty\\
&= \frac{\pi}{4} \left (Q_\infty^2 -\sum_{a=1}^N Q_a^2\right)
\end{align}
This is equal to half the difference in entropy between the initial and final configurations. This is different from the value of the Euclidean action for Brill's instanton, which is equal to the exact difference in entropy between a single $AdS_2\times S_2$ throat and multiple completely disconnected $AdS_2\times S_2$ throats. In some sense, it can be seen as natural that the probability amplitude for our instanton is half that for the Brill instanton, as the instanton in this paper, unlike Brill's, connects a single $AdS_2\times S_2$ throat to a configuration that branches into several $AdS_2\times S_2$ throats at the bottom of the single throat, but remains connected at the top. Therefore, it can be said in some sense that only ``half'' of the original space becomes disconnected in the transition represented by the instanton, so that the corresponding tunneling amplitude is also half that of the process which gives a set of completely disconnected spacetimes.

\section{Conclusion}\label{sec-Conc}

We have found an instanton that can be interpreted as a tunneling process between a single-centered black hole solution of charge $Q_\infty$ to a multi-centered black hole solution of charges $Q_1,\dots,Q_N$ such that $Q_\infty = Q_1 + \dots + Q_N$. The amplitude for the tunneling process is equal to half the difference in entropy between the initial and final configurations. The black holes that we are considering are contained within an encapsulating $AdS_2\times S_2$ throat. Thus we may consider them as black holes in a BR universe, or alternatively, as an $AdS_2\times S_2$ throat that divides into multiple branches as we move deeper into the throat. In the latter interpretation, our instanton describes the splitting of the throat of an ERN black hole by quantum tunneling. 

Ultimately, it would be desirable to find an instanton that describes the complete splitting of an ERN black hole into two or more ERN black holes separated by finite coordinate distances in an asymptotically flat space. Our instanton suggests that this process is not forbidden, as there is at least a finite probability for the throat of an ERN black hole to split into multiple throats.

\section{Acknowledgements}

I would like to thank Frederik Denef for helpful discussions on this work, and Andrei Linde for pointing me towards some useful references. This project was funded in part by a Research Assistantship from Harvard's Center for the Fundamental Laws of Nature.


\begin{thebibliography}{14}
\bibitem{Wheeler} J.~A.~Wheeler, Phys. Rev. 97, 511 (1955).
\bibitem{Garfinkle} D.~Garfinkle and A.~Strominger, Phys. Lett. B 256, 146 (1991).
\bibitem{Brill} D.~Brill, Phys. Rev. D46 (1992), 1560, arXiv:hep-th/9202037.
\bibitem{AdSFrag} J.~Maldacena, J.~Michelson, and A.~Strominger, JHEP9902, (1999), 011, arXiv:hep-th/9812073.
\bibitem{SUSYCC} R.~Kallosh, A.~Linde, T.~Ortin, A.~Peet, and A.~Van Proeyen, Phys. Rev. D46 (1992), 5278, arXiv:hep-th/9205027.
\bibitem{BertRob} T.~Levi-Civita, R. C. Acad. Lincei (5) \textbf{26}, 519 (1917), B.~Bertotti, Phys. Rev. \textbf{116}, 1331 (1959), I.~Robinson, Bull. Akad. Polon. \textbf{7}, 351 (1959).
\bibitem{GibbonsHartle} G.~W.~Gibbons and J.~B.~Hartle, Phys. Rev. D42 (1990), 2458.
\bibitem{Hayward} G.~Hayward, Phys. Rev. D47 (1993), 3275.
\bibitem{HawkingIsrael} S.~Hawking and W.~Israel, \textit{General Relativity: An Einstein Centenary Survey}, Cambridge University Press. 
\bibitem{PiolineTroost} B.~Pioline and J.~Troost, JHEP0503, (2005), 043, arXiv:hep-th/0501169.
\end{thebibliography}
\end{document}